\documentclass[prb,twocolumn,showpacs]{revtex4}
\usepackage{bm}

\usepackage{epsfig}
\usepackage{ulem}
\newcommand{\nix}[1]{}
\usepackage{color}
\usepackage{graphics}

\begin{document}

\title{ Interplay of  spin and orbital magnetogyrotropic 
photogalvanic effects in InSb/AlInSb quantum well structures}
\author{ S.~Stachel,$^1$ P.~Olbrich,$^1$ C.~Zoth,$^1$ U.~Hagner,$^1$
T.~Stangl,$^1$ C.~Karl,$^1$ P.~Lutz,$^1$ V.V.~Bel'kov,$^{2}$ 
 S.K.~Clowes,$^{3}$ T.~Ashley,$^{4}$ 
A.M. Gilbertson,$^{5}$ and S.D.~Ganichev$^{1}$
}

\affiliation{$^1$Terahertz Center, University of Regensburg, 93040
Regensburg, Germany} 
\affiliation{$^2$Ioffe Physical-Technical
Institute, Russian Academy of Sciences, 194021 St.\,Petersburg,
Russia} 
\affiliation{$^3$Advanced Technology Institute and SEPNet, University
of Surrey, Surrey GU2 7XH, UK} 
\affiliation{$^4$ School of Engineering, University of Warwick, Coventry CV4 7AL, UK}
\affiliation{$^5$ Blackett Laboratory, Imperial College, London SW7 2BZ, UK}

\begin{abstract}

We report on the observation of linear and circular 
magnetogyrotropic photogalvanic effects in InSb/AlInSb quantum well structures. 
We show that intraband (Drude-like) absorption of terahertz radiation in 
the heterostructures causes a \textit{dc} electric current in the presence 
of an in-plane magnetic field. The photocurrent behavior upon variation 
of the magnetic field strength, temperature and wavelength is studied. 
We show that at moderate magnetic fields the photocurrent exhibits a typical 
linear  field dependence. At high magnetic fields, however, it becomes nonlinear 
and inverses its sign. The experimental results are analyzed in terms of 
the microscopic models based on 
asymmetric relaxation of carriers in the momentum space. 
We demonstrate that the observed nonlinearity 
of the photocurrent is caused by the large Zeeman spin splitting in 
InSb/AlInSb structures and an interplay of the spin-related and 
spin-independent roots of the magnetogyrotropic photogalvanic effect.
\end{abstract}
\pacs{73.21.Fg, 72.25.Fe, 78.67.De, 73.63.Hs}

\maketitle

\section{Introduction}

Indium antimonide based quantum wells (QWs) have attracted growing 
attention for high-speed transistors~\cite{Ashley2007}, quantum computing~\cite{Khodaparast2004,Litvinenko2006} and 
infrared lasers~\cite{andreevapl2001}.
This novel material is the subject of numerous experimental studies of transport, 
optical, magneto-optical and spin-related 
phenomena~\cite{Ashley2007,Khodaparast2004,Litvinenko2006,andreevapl2001,Santos3,Santos1,Santos2,Orr2007,Orr2008,Gilbertson,Nedniyom2009,Pooley2010,Leontiadou,Santos2011}. 
The characteristics driving  the interest in this novel narrow gap
material are the high carrier mobility, small effective masses, large Land\'{e} $g^*$-factor, 
possibility of the mesoscopic spin-dependent ballistic transport and  a strong spin-orbit coupling.  
The latter gives rise to a number of optoelectronic effects such as, e.g.,
terahertz photoconductivity~\cite{vasiliev} 
and  the circular photogalvanic effect~\cite{Ivchenko08,PhysicaE02,PRB2003,PRB03sge,Bieler05,Yang06,Cho07}
recently observed in InSb QWs~\cite{Khodaparast}. 
Investigation of photogalvanic effects in the presence 
of a magnetic field  should provide further access to nonequilibrium processes in 
low-dimensional structures yielding information of such details as the anisotropy of the
band spin-splitting, processes of momentum and energy relaxation,
symmetry properties and the Zeeman spin splitting, 
for review see~\cite{Ivchenko08,Ganichevbook,Belkov2008}.

Here we report on the observation and detailed study of the 
magneto-gyrotropic photogalvanic effects (MPGE)~\cite{Belkov2008,Belkov2005}
in $n$-doped  InSb/AlInSb QWs induced by 
terahertz (THz) radiation. 
We discuss both the linear magnetogyrotropic photogalvanic effect (LMPGE), which can be induced by linearly polarized or 
unpolarized radiation, as well as the circular magnetogyrotropic photogalvanic effect (CMPGE), which results in the light 
helicity dependent photocurrent and reverses its direction upon 
switching the sign of the circular polarization. 
We show that in InSb/AlInSb QWs the narrow energy gap and the strong 
spin-orbit coupling combined 
with the large Land\'{e} $g^*$-factor result in a photocurrent
orders of magnitude larger than that reported for GaAs- 
and InAs-based QWs, for review see~\cite{Belkov2008}.
Moreover, in contrast to previous studies, the observed photocurrent 
exhibits a peculiar magnetic field dependence: 
while for moderate magnetic fields ($< 1$~T) the LMPGE current has a typical 
linear dependence on magnetic field $\bm B$, 
at higher magnetic fields  it becomes nonlinear and reverses its sign. 
By contrast, the CMPGE  remains linear in the whole range of  investigated magnetic fields. 
The experimental results are analyzed in terms of 
spin~\cite{Nature02,Nature06,Lechner11} 
and orbital~\cite{Lechner11,Tarasenko_orbital,Tarasenko_orbitalMPGE2} 
microscopic models of the magneto-gyrotropic photogalvanic effect based on the asymmetry 
of the relaxation of carriers in the  momentum space. 
We demonstrate that specific magnetic field dependences observed for the LMPGE
are due to the nonlinear Zeeman spin-splitting in InSb/AlInSb QWs which is 
enhanced by the electron-electron exchange interaction   and   
causes a nonlinear increase of the spin-related MPGE.

\section{Samples and experimental techniques}
\label{technique}

\begin{figure}[h]
\includegraphics[width=\linewidth]{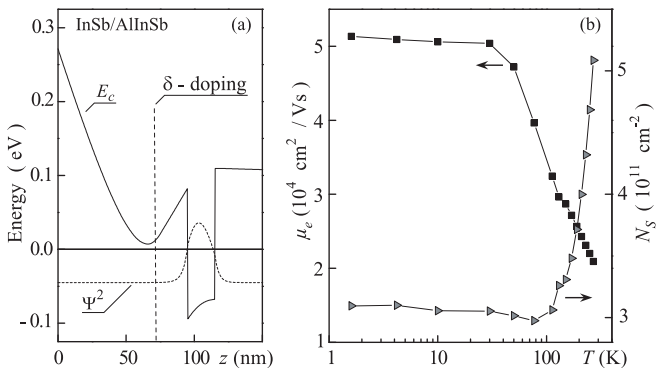}
\caption{a) Conduction band profile and electron wavefunction of  QW structure
with $L_W = $ 20~nm calculated
within a self consistent Schr{\"o}dinger-Poisson model~\cite{Leontiadou}. 
b) Temperature dependences of mobility $\mu_e$ and carrier density $N_{s}$ 
obtained by the low-field Hall measurements in 
20 nm QW sample.
}
\label{fig1}
\end{figure}

\begin{figure}[h]
\includegraphics[width=0.9\linewidth]{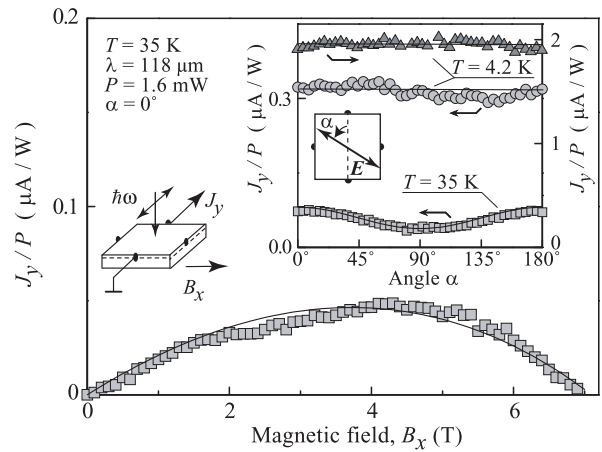}
\caption{Magnetic field dependence of $J_{y}/P$ for $\lambda =
118$~$\mu$m and $T = 35$~K. Lines are fit after Eq.~(\ref{fit}).
The left inset shows the experimental geometry.
The right inset shows the  photocurrent as a function of the azimuth 
angle $\alpha$ measured for $T = 4.2$ and $35$~K at fixed $B_{x}$ = +5~T. Triangle symbols correspond to 30~nm, circle and squared symbols to 20~nm QW structures.
}
\label{fig2}
\end{figure}

We investigated two \textit{n}-type InSb/AlInSb single quantum well 
structures grown by molecular beam epitaxy onto semi-insulating nominally (001)-oriented 
GaAs substrate. A QW of width, $L_W$, 
is confined on each side by an InAlSb barrier, 
with a Te modulation doped layer 20~nm above the QW (ME1833 and ME2507 with $L_W =$~20~nm and 30~nm, respectively)~\cite{Leontiadou,Gilbertson}.
The calculated conduction band profile, electron wave function and doping position of the 20~nm QW are shown
in Fig.~\ref{fig1}(a). The data are obtained by a self-consistent solution of the 
Schr{\"o}dinger and Poisson equations~\cite{Leontiadou}. 
The QW with $L_W =$~20~nm (30~nm) width contains a 
two dimensional electron gas with the carrier 
density of $N_{s}\approx 3 \times 10^{11}$~cm$^{-2}$ ($5 \times 10^{11}$~cm$^{-2}$)
and the mobility of $\mu_e \approx 5 \times 10^{4}$~cm$^2/$~V$\cdot$s ($15 \times 10^{4}$~cm$^2/$~V$\cdot$s) 
for $T$ below 77~K. 
The temperature dependence of  $\mu_e$ and $N_{s}$ measured in the 20~nm QW 
structure by low-field Hall effect are shown in  Fig.~\ref{fig1}(b). 
The samples have  square shape and two pairs of ohmic contacts on 
opposite side of the edges (see inset in Fig.~\ref{fig2}) oriented along  
$x\parallel[1\bar{1}0]$ and
$y\parallel[110]$. 
The photocurrents have been investigated in the temperature range of
$T = 4.2 $ to 270~K using an optical cryostat with a superconducting
magnet.  The external
magnetic field $\bm B$ up to $\pm$7~T has been applied 
parallel to the interface plane along $x$-direction. 

To generate photogalvanic currents we  applied a continuous-wave ($cw$) 
and pulsed molecular lasers optically pumped by CO$_{2}$ lasers.
For low power $cw$ radiation we used
a CH$_3$OH  laser operating at wavelength $\lambda = 118\,\mu$m (frequency $f = 2.5$~THz)
with a power $P \approx 2$\,mW at the sample position. 
The radiation was modulated at
120~Hz, allowing the detection of the photoresponse  by the 
standard lock-in technique.
High power radiation is obtained by a pulsed NH$_3$ laser 
optically pumped by a transversely excited
atmosphere TEA-CO$_{2}$ laser and operating at wavelengths 
$\lambda \, = \, 90.5$, 148 or 280\,$\mu$m
(frequencies $f \, = \, 3.3$, 2 and 1.1\,THz, respectively).
More details on the system can be found
in~\cite{JETP1982,PRL93,PRL95,Schneider2004,PhysicaB99tun}.
Here, we used  single pulses with a duration
of about 100~ns, peak power of $P \approx $ 5~kW, and a
repetition rate of 1~Hz.
The photocurrents in unbiased structures are measured via 
the voltage drop across a 50~$\Omega$ load
resistor with a storage oscilloscope.
The radiation power of $cw$ and pulsed radiation has 
been controlled by a pyroelectric detector, calorimeter and
THz photon drag detector~\cite{Ganichev84p20}, respectively. 
A typical spot diameter is from 1 to 3~mm.
The beam has an almost Gaussian form,
which is measured by a pyroelectric 
camera~\cite{ch1Ziemann2000p3843}.

All experiments are performed at normal incidence of light. 
Photocurrents are measured perpendicularly  ($J_{y}$) and parallel ($J_{x}$) 
to the applied magnetic field ($B_{x})$, referred to as tranverse and longitudinal photocurrents, respectively.
Our lasers emit linearly polarized radiation with 
the electric field vector of the THz radiation oriented 
along the $y$-axis. 
In order to rotate the electric field vector $\bm E$ 
by the angle $\alpha$ ($\alpha=0^\circ$, $\bm E\parallel y$) we used a $\lambda$/2 plate. 
To excite the circular photocurrent we changed the
radiation helicity, $P_{circ}$, by rotating the $\lambda$/4 plate by the angle $\varphi$ 
between the initial linear polarization of the 
laser light and the plate optical axis.
In this way, the helicity of the incident light can be
varied from $-1$ (left-handed circular, $\sigma_-$) to $+1$
(right-handed circular, $\sigma_+$) according to $P_{circ} =
\sin{2 \varphi}$.

\section{Results}
\label{results}

\subsection{Photocurrent induced by  linearly polarized/unpolarized radiation}

We shall start by describing the results obtained by irradiating the sample
with linearly polarized radiation which may result only in signals 
due to the LMPGE and excludes the CMPGE.
The  magnetic field induced photocurrent
is studied by applying an in-plane magnetic field $B_{x}$. 
The observed signal varies with magnetic field strength and 
its sign depends on the magnetic field direction.
While for the 30~nm QW sample no signal is detected at zero magnetic field, in samples with 
$L_W = 20$~nm QW we observed a signal at $B_{x}$ = 0.
The origin of this magnetically independent signal~\cite{symmetry}
is not within the scope of this paper and will be 
discussed elsewhere. In the following, we eliminate this 
contribution by taking $J(|B|)$ as
\begin{equation}
\label{MPGE} J_{y}(|B|) = [J(B_x > 0) - J(B_x < 0)]/2
\end{equation}
so that only magnetic field dependent effects remain.

Transverse photocurrent $J_y(|B|)$ excited by the linearly polarized 
radiation of low power \textit{cw} laser 
is shown in Fig.~\ref{fig2} as a function of the magnetic field $B_{x}$ 
and in the right inset in Fig.~\ref{fig2} as a function of 
the azimuth angle $\alpha$ for a fixed magnetic field $B_{x}$ = +5~T.
While at low temperatures the transverse photocurrent $J_y$
comes almost all from the polarization independent offset, at 
higher temperature we 
observed a  variation of the photocurrent with rotation of linear polarization 
($J_y = J_1 + J_2\cos(2\alpha)$). 
In the longitudinal configuration we detected only the polarization dependent photocurrent 
$J_x = J_3\sin(2\alpha)$ which, 
like transversal partial current $J_2\cos(2\alpha)$, contributes at higher temperatures only.
The experiment reveals that, in particularly at low temperatures, 
the polarization dependent photocurrent contributions,
$J_2\cos(2\alpha)$ and $J_3\sin(2\alpha)$, in our samples are substantially smaller than $J_1$. 
Thus in the following we focus on the polarization 
independent photocurrent $J_y$ observed in the transverse geometry.

\begin{figure}[h]
\includegraphics[width=0.9\linewidth]{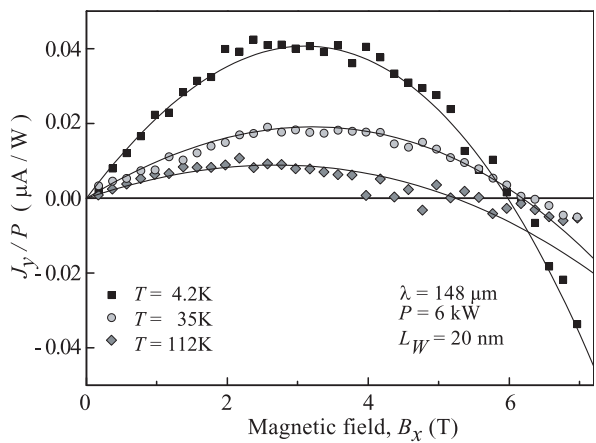}
\caption{Magnetic field dependence of $J_{y}/P$ for $\lambda =
148$~$\mu$m and different temperatures. Lines are fit after Eq.~(\ref{fit}). 
}\label{fig3}
\end{figure}

\begin{figure}[h]
\includegraphics[width=0.9\linewidth]{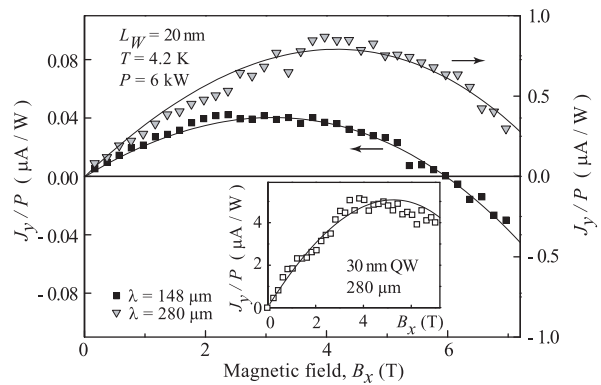}
\caption{Dependence of the LMPGE on the magnetic field 
at $T = 4.2$~K for wavelengths of $\lambda = 148$ and $280~\mu$m obtained for 20~nm QW 
structure. The inset shows the LMPGE for 30~nm QW sample.
Lines are fit after Eq.~(\ref{fit}).
}\label{fig4}
\end{figure}

The most striking observation comes from the 
investigation of the magnetic field
dependence of the photocurrent. 
The general behavior of the photocurrent 
is that  the magnitude of $J_{y}$ is 
proportional to $B_{x}$ for low magnetic fields only.
At higher fields, however, the signal becomes nonlinear: 
with increasing $B_{x}$ the sign of $dJ_{y}/dB_{x}$ 
changes and, finally, the signal vanishes and for some conditions even reverses its sign. 
Figure~\ref{fig2} shows such a magnetic field dependence measured
applying low power radiation of the $cw$  laser with $P \approx 1.6$~mW.
In Figs.~\ref{fig3} and ~\ref{fig4} we plotted the magnetic field dependence of $J_{y}$ excited by the high power radiation of the pulsed laser. 
The data obtained for a fixed wavelength of $\lambda$ = 148~$\mu$m and various temperatures (Fig.~\ref{fig3}) and for fixed temperature of 4.2~K but several wavelengths 
(Fig.~\ref{fig4}).
Figure~\ref{fig3} depicts that increasing the temperature reduces 
the magnitude of the photocurrent , whereas the magnetic field $B_x \approx 6.2$~T at which the zero-crossing occurs remains almost unchanged.
Measuring the temperature dependence for both low power and high power 
excitations we obtained that for $T< 8$~K the photocurrent is constant at fixed magnetic field and 
at higher temperatures rapidly decays showing close to 
$J \propto 1/T$ behavior (not shown).
In the case of fixed temperature but increasing wavelength (see Fig.~\ref{fig4})
the magnitude of the photocurrent 
increases  and the zero-crossover is shifted to higher magnetic fields. 
Finally we note, that sweeping the magnetic field from negative to positive and back 
we did not observe a hysteresis.

\begin{figure}[h]
\includegraphics[width=0.75\linewidth]{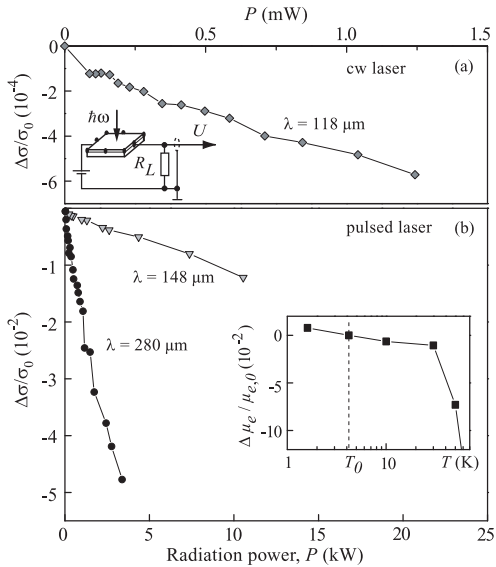}
\caption{  Relative change in conductivity, 
$\Delta\sigma/\sigma_0 = (\sigma_i-\sigma_0)/\sigma_0$,  
in QW structure with $L_W = 20$~nm 
measured  versus radiation power $P$ at $T = 4.2$~K and $B=0$.
The ratio of conductivity under illumination, $\sigma_i$, 
and dark conductivity, $\sigma_0$,
is determined from the photoconductive
signals measured in the circuit sketched in the inset of the upper plate. 
(a) Photoconductive signal  measured applying  $cw$ radiation 
with  wavelength  $\lambda = 118$~$\mu$m. 
(b) $\Delta\sigma/\sigma_0$ measured 
applying   pulsed laser radiation 
with $\lambda = 148 $ and $ 280$~$\mu$m. 
The inset shows a section of the temperature dependence of
the relative mobility $\Delta \mu_e / \mu_{e,0}$, where $\mu_{e,0}$ is the mobility at $T_0 = 4.2$~K. 
}
\label{fig5}
\end{figure}

Our experiments demonstrate that the photocurrent is dominated by a photocurrent 
contribution which is insensitive to the radiation polarization. Earlier studies of 
magneto-photocurrents demonstrated that such polarization independent photocurrents are 
caused by the radiation induced electron gas heating followed by the 
scattering asymmetry in $\bm{k}$-space, see Refs.~[\onlinecite{Belkov2008,Belkov2005,Nature06}].
In order to characterize the electron gas heating in our structures we 
investigated the THz-photoconductivity applying the same wavelengths and powers.
Figure~\ref{fig5}(a) shows the photoconductive signal excited by $cw$ THz laser
as a function of radiation power.
The observed decrease of the structure conductivity with increasing 
THz radiation (negative photoconductivity) provides 
the  evidence for the electron gas heating. Indeed, Hall 
measurements, see  Fig.~\ref{fig1}(b) and the inset in Fig.~\ref{fig5}(b), 
show that a rise of temperature results in the decrease of mobility 
and, consequently, in the lowering of conductivity.
The data for pulsed excitation,  presented in Fig.~\ref{fig5}(b), demonstrate
that an increase of the radiation power by 
about six orders of magnitude results in a change of the relative 
photoconductivity $|\Delta \sigma / \sigma_0|$ by two orders of magnitude.
We attribute the observed nonlinearity of the photoconductive response to
nonlinear energy losses in InSb QWs at low temperature, 
which, consequently, cause a strongly nonlinear dependence of the
electron temperature on the absorbed energy~\cite{Ganichevbook}.
Comparison of the data obtained at  $\lambda=148$ and 280~$\mu$m
demonstrates essentially stronger electron gas heating at a longer wavelength. 
This observation is in a good agreement with the frequency dependence of the
Drude-like absorption. Figure~\ref{fig5}(b) shows that for radiation power of several kilowatts 
relative photoconductivity achieves values as high as 10$^{-2}$ to 10$^{-1}$.
Comparison of these values with the mobility data [see the inset in Fig.~\ref{fig5}(b)] shows
that pulsed THz radiation used here can heat up the 
electron gas by tens kelvin\,\cite{heating}.
%

\subsection{Photocurrent induced by circularly polarized radiation}

\begin{figure}[h]
\includegraphics[width=0.9\linewidth]{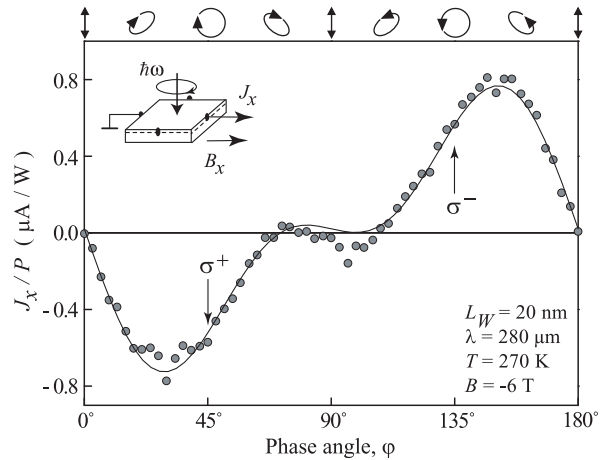}
\caption{Helicity dependence of the photocurrent $J_{x}$ measured for
$B_{x}= -6$~T and  $\lambda = 280$~$\mu$m with subtracted offset $\xi$. The inset
shows the experimental geometry. The ellipses on top illustrate 
the polarization states for various
$\varphi$.
}\label{fig6}
\end{figure}

\begin{figure}[h]
\includegraphics[width=0.9\linewidth]{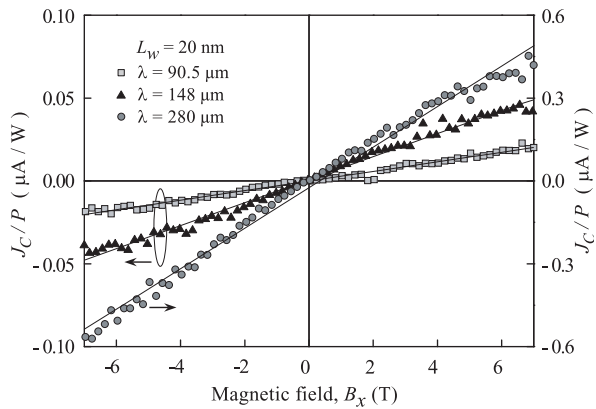}
\caption{Magnetic field dependence of $J_{x}/P$ for wavelengths of $\lambda = 90.5, 148$ 
and $280$~$\mu$m at $T = 270$~K. 
}
\label{fig7}
\end{figure}

We will now describe the results for irradiation 
with circularly (elliptically) polarized light
which is obtained using a $\lambda/4$ plate. 
The ellipses on top of Fig~\ref{fig6} illustrate 
the polarization states for various angles $\varphi$.
The resulting polarization state is given by
the Stokes parameters~\cite{Stokes} 
$S_1 = \cos^2 (2\varphi)$ and $S_2 = \sin (4\varphi) /2$,  
describing the degree of linear polarization, 
and $S_3 \equiv P_{\rm circ}$.
The photocurrent detected in the transverse geometry is well described by
$J_y = J_1 + (J_2/2) \cos (4\varphi)$. The  
photocurrent consists of polarization independent contribution, $J_1$,
and a contribution that is proportional to the degree of linear polarization, i.e. 
just the same as discussed in the previous section. 
In the longitudinal geometry ($J_{x} \parallel B_{x}$), however,
we observed a new contribution to the photocurrent. 
It manifests itself in the helicity dependence of the signal.
The dependence of the photocurrent $J_{x}$ on $\varphi$ is shown in Fig.~\ref{fig6}.
It is well described by $J_x(\varphi) = (J_3/2) \sin(4\varphi) + J_C \sin(2\varphi) +  \xi$.
Here the first term is again just the contribution proportional to $J_3$ 
in the described above experiments with linearly polarized radiation. 
It reflects the degree of linear polarization and vanishes for 
circularly polarized light. The second term is proportional to 
the radiation helicity $P_{circ}$. 
This circular photocurrent 
changes its sign by switching the light helicity from $-1$ 
to $+1$. 
Note that the observed offset $\xi$ is much smaller than $J_3$ and $J_C$
and is subtracted from the data of Fig.~\ref{fig6}.
We will focus on circular photocurrent in the longitudinal geometry, thus we can extinguish all other possible effects by
\begin{equation}
\label{CMPGE} J_C = [J_{x}(\sigma^{+}) - J_{x}(\sigma^{-})]/2 .
\end{equation}
Figure~\ref{fig7} shows the magnetic field dependence of the circular 
photocurrent  $J_C$ measured in the 20~nm QW structure for 
different wavelengths. Similarly to the 
photocurrent induced by linearly polarized radiation
its magnitude normalized by the radiation power substantially  increases for longer 
wavelengths: the fact which can also naturally be attributed to 
the increase of the Drude-absorption.
However, unlike the photocurrent induced by linearly polarized radiation (Fig.~\ref{fig2}~-~\ref{fig4}), the circular 
photocurrent remains proportional to the magnetic field $B_{x}$ up to the highest 
field applied, $|B| = 7$~T. The same behavior has been observed in the 30~nm structure (not shown).

\section{Discussion}
\label{discussion}

All our observations at low magnetic field exhibit the recognized 
MPGE behavior, which by definition is a
magnetic field induced photocurrent related to the gyrotropic
symmetry of the system~\cite{Belkov2005}. In particular, the observed linear coupling 
to the magnetic field, the in-plane anisotropy of the photocurrent, as 
well as the polarization dependences all follow the symmetry arguments for the MPGE. 
The current perpendicular to the magnetic field is dominated by the polarization 
independent contribution (LMPGE) and is therefore driven 
by relaxation processes~\cite{Nature06,PRBSiGe}. 
At the same time the longitudinal magnetic field induced 
photocurrent excited 
by circularly polarized radiation is solely governed by 
the photon angular momentum (CMPGE)~\cite{Nature02}. 
Comparison of our data on the magnetic field induced
photocurrents with that reported earlier for GaAs- and 
InAs-based QW structures (for review see~\cite{Belkov2008}) shows that 
it is much stronger in InSb QWs  by at least two orders of magnitude.
We note that for the 30~nm QWs 
compared to our 20~nm QW we detected ten times 
larger photoresponses, see the inset in Fig.~\ref{fig2} 
and the data for $\lambda = 280$~$\mu$m  in Fig.~\ref{fig4}.
While the general features of our signals are in agreement with previous results for 
III-V QWs, the magnetic field dependences of the LMPGE and the CMPGE in InSb-based QWs 
have a contradictory behavior: the LMPGE is nonlinear and the CMPGE is linear. 
We will now discuss separately the LMPGE 
and CMPGE in terms of the interplay between the spin and 
orbital (non-spin) related relaxation processes. We will show that 
this interplay results in the surprising magnetic field behavior.

\begin{figure}[h]
\includegraphics[width=\linewidth]{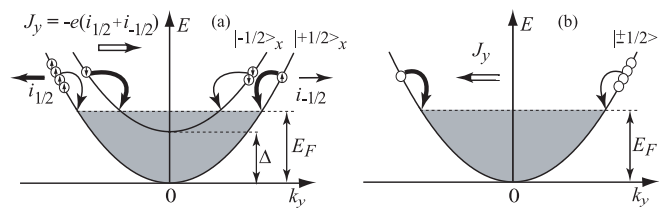}
\caption{Models of magneto-gyrotropic photogalvanic currents: (a) -
spin-dependent LMPGE; (b) - orbital LMPGE.}
\label{fig9}
\end{figure}

\subsection{Linear MPGE}

The spin-related origin of the LMPGE is
a consequence of the electron gas heating followed by spin 
dependent scattering~\cite{Belkov2008,Nature06,PRBSiGe}. The latter is due to 
the spin-orbit interaction in gyrotropic media, such as 
InSb- and GaAs-based low-dimensional structures, which yields a 
scattering matrix element being proportional 
to $\left[ \bm{\sigma} \times \left( \bm{k} +
\bm{k^\prime} \right) \right]$. Here $\bm{k}$ and $\bm{k^\prime}$
are the initial and the scattered wave vectors and $\bm{\sigma}$
is the vector composed of the Pauli matrices and only structural inversion 
asymmetry is assumed. This spin dependent scattering results in an
asymmetric relaxation of the hot electrons shown by the different
thickness of the arrows in Fig.~\ref{fig9}(a) and causes
oppositely directed electron fluxes $\bm{i}_{\pm1/2}$ in the spin
subbands. Consequently, a spin current, defined as difference
between the fluxes, is given by $\bm{J}_{s} = 1/2 (\bm{i}_{+1/2} - \bm{i}_{-1/2})$.
At nonzero magnetic field, e.g., $B_x$, the Zeeman effect causes an 
equilibrium spin polarization parallel to the magnetic field 
and the fluxes become unbalanced due to the unequal equilibrium 
population of the spin subbands.  The average electron spin $s$ 
is equal to
\begin{equation}
s = \frac12 \frac{N_{+1/2}-N_{-1/2}}{N_{+1/2}+N_{-1/2}} \:.
\end{equation}
Such an imbalance results in a net electric current $\bm{j}_{\rm{spin}}$
given by the sum of the fluxes, 
$\bm{j}_{\rm spin} = - e (\bm{i}_{+1/2} + \bm{i}_{-1/2})$, where $-e$ is the electron charge. 
Assuming that the fluxes  $\bm{i}_{\pm1/2}$ are proportional to 
the carrier densities in the spin subbands $N_{\pm 1/2}$, one obtains
\begin{equation}\label{current1}
\bm{j}_{\rm spin} = - 4 e s \bm{J}_{s} \:.
\end{equation}
We note that while
in the theoretical consideration the current density $\bm{j}$ is
used, in the experiments the electric current $\bm{J}$ is measured
which is proportional to the current density $\bm{j}$.

At low magnetic fields with the Fermi energy $E_F$ larger than the energy of the Zeeman spin splitting, 
$\bm s$ is a linear function of magnetic field $\bm B$ and is given by
\begin{equation}
\label{S_low} 
\bm{s} = - \frac{\Delta}{4E_F} \frac{\bm{B}}{B} \:,
\end{equation}
where \mbox{$\Delta = g^* \mu_B B$} is the energy of the Zeeman spin splitting
and $\mu_B$ is the Bohr magneton. 
However, in the high field limit for $|\Delta| > 2E_F$
one of the spin subbands 
will be completely depopulated. Obviously in this case the average 
spin $s = \pm 1/2$ and  $J_{y}$ saturates.

The behavior of the spin-dependent LMPGE, $J_{y} \propto s$, over all 
magnetic fields can be obtained 
taking into account that in thermal equilibrium, 
the densities $N_{\pm 1/2}$ are determined by 
\begin{equation}
N_{\pm 1/2} \propto \sum_{\bm{k}} \left[ \exp\left( \frac{\varepsilon_{\bm{k}} \pm \Delta/2 - \mu }{k_B T_e} \right) + 1 \right]^{-1}  \:,
\end{equation}

where $\varepsilon_{\bm{k}}=\hbar^2 k^2/(2m^*)$ is the kinetic energy, $m^*$ is the effective mass,
$\mu$ is the chemical potential, $k_B$ is the Boltzmann constant, and $T_e$ is the electron temperature. 
Effects on $N_{\pm 1/2}$ due to nonparabolicity of the subbands~\cite{Litvinenko2006} will be weak compared to the Boltzmann redistribution from the Zeeman spin splitting, and are therefore ignored.
Straightforward summation over the wave vector $\bm{k}$ yields
\begin{equation}\label{eq3}
s = \frac12 \, \frac{\ln\left\{ \left[ 1 + \exp \left( \frac{\mu-\Delta/2}{k_BT_e} \right) \right] / 
\left[ 1 + \exp \left( \frac{\mu+\Delta/2}{k_BT_e} \right)  \right]  \right\}}
{\ln\left\{ \left[ 1 + \exp \left( \frac{\mu-\Delta/2}{k_BT_e} \right) \right] 
\times \left[ 1 + \exp \left( \frac{\mu+\Delta/2}{k_BT_e} \right)  \right]  \right\}} \:.
\end{equation}
Equation~(\ref{eq3}) describes the average spin of two-dimensional carriers in an 
external magnetic field for a fixed chemical potential $\mu$. If, instead, the carrier 
density $N_s = N_{+1/2}+N_{-1/2}$
like in our case 
is fixed, Eq.~(\ref{eq3}) should be supplemented with the following 
equation for the chemical potential  
%
%
\begin{eqnarray}
\mu = k_B T_e  \ln \left[ \sqrt{ \exp\left( \frac{2\pi N_s \hbar^2}{m^* k_B T_e} \right) 
+ \cosh^2 \left( \frac{\Delta}{2k_BT_e}\right) -1 } 
\right. \: 
\\
\left. 
- \cosh \left( \frac{\Delta}{2k_BT_e}\right) \right] \:. \nonumber
\end{eqnarray}

\begin{figure}[h]
\includegraphics[width=\linewidth]{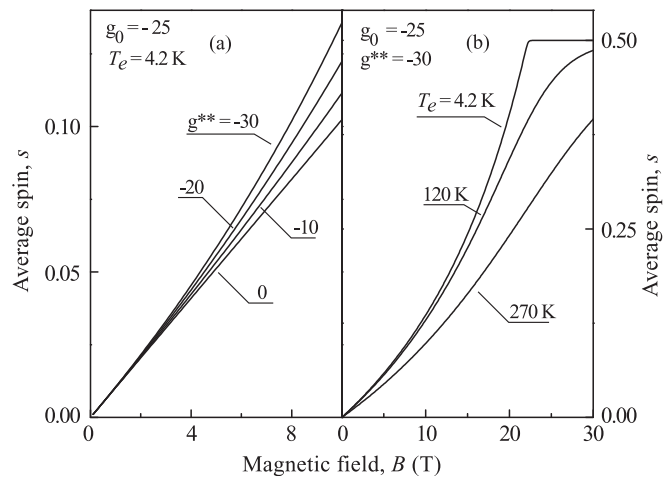}
\caption{Average spin in 20~nm QW structures obtained by self-consistent calculations of Eqs.~(\ref{eq3}) 
and~(\ref{Delta_exc}) as a function of the magnetic field. For calculation we used $g_0 = -25$ and an
effective mass $m^*=0.02~m_0$.
Average spin calculated for  (a) fixed temperature but for various values of the exchange 
interaction given by the parameter $g^{**}$ indicated by numbers next to the curves; (b) fixed exchange interaction, $g^{**} = -30$
but various electron temperatures $T_e$.
} \label{fig8}
\end{figure}

The magnetic field dependence of $s$ given by Eq.~(\ref{eq3}) is nonlinear saturating at $|s|=1/2$. However, the deviation from linear dependence for the degenerate electron gas occurs at rather high magnetic fields when the average spin projection is close to $\pm 1/2$. Therefore, we suggest that other effects resulting in a nonlinear magnetic field dependence of the electron spin are responsible for the observed reversal of the electric current with the field increase. As a possible origin of this effect we consider {\it exchange interaction} between electrons which is known to lead to a nonlinearity of the Zeeman splitting on the external magnetic field at moderate fields~\cite{Nedniyom2009,Santos2011}.
In this case the effective $g^*$-factor besides $g_0$, Land\'{e} factor at $B=0$,  
contains a contribution linear in the spin polarization 
\begin{equation}\label{Delta_exc}
g^* = g_0 + 2 |s| g^{**} \:,\;\;\; \Delta = ( g_0 + 2 |s| g^{**}) \mu_B B \:,
\end{equation}
where   $2 |s| g^{**}$
is the contribution to $g^*$-factor caused by the exchange interaction.   
Equations~(\ref{eq3}) and~(\ref{Delta_exc}) supplement each other and 
are to be solved together.  The calculated magnetic field behavior of the average spin 
and, consequently, the photocurrent ($J_y \propto s$), 
is plotted in Fig.~\ref{fig8}. For calculation we used $g_0 = -25$ and an
effective mass $m^*=0.02~m_0$ determined by magneto-transport 
experiments~\cite{Nedniyom2009} and cyclotron resonance data (not shown), respectively.
Figure~\ref{fig8}(a) shows the average spin calculated for 20~nm QW 
structures at fixed temperature but for various values of the exchange 
interaction given by the parameter $g^{**}$.
At low temperatures, for $g^{**} = 0$ and $|\Delta| \leq 2E_F$,
we obtain a linear dependence 
of the average electron spin on the Zeeman splitting following the well 
known behavior described by Eq.~(\ref{S_low}).
The exchange interaction results in a superlinear magnetic field dependence of $s(B)$ 
so that for $g^{**} = - 30$ the average spin is substantially enhanced already at a magnetic 
field of several Tesla. Finally for  $|\Delta| > 2E_F$ 
one of the spin subbands 
will be completely depopulated and  $|s| = 1/2$.
Note that $g^{**} = - 30$ is obtained in InSb QWs similar to our structures
by magneto-transport measurements~\cite{Nedniyom2009}.
 Using this value we 
calculated how electron temperature influences the average spin. 
The results are plotted in Fig.~\ref{fig8}(b). The data shows that 
an increase of the temperature results in a decrease of spin polarization and 
reduces the nonlinearity. However, for magnetic fields below 7~T and temperatures 
below $\approx 130$~K, used in experiments here, $s$ remains nearly unchanged by the temperature.

\begin{figure}[h]
\includegraphics[width=0.45\linewidth]{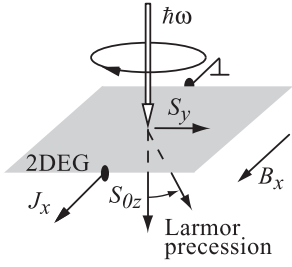}
\caption{Model for the spin-related CMPGE. The excitation  with
circularly polarized light yields a spin orientation $S_{0z}$. An
in-plane component $S_{y}$ of the nonequilibrium spin is generated
by the Larmor precession.}\label{fig10}
\end{figure}

While spin mediated relaxation can produce a non-linear signal it 
cannot cause the observed sign reversal of the photocurrent. 
Thus we consider another known mechanism of the LMPGE based on an asymmetric 
relaxation due to the Lorentz force acting on heated carriers~\cite{Lechner11,Tarasenko_orbital,Tarasenko_orbitalMPGE2}
which may provide an additional contribution to the total photocurrent.
The effect is illustrated in Fig.~\ref{fig9}(b). Similar to the
spin-related MPGE, the current stems from the asymmetric energy relaxation 
of the hot electrons. 
Now, however, this asymmetry is caused
by the scattering correction being linear in the wavevector $\bm{k}$ and in the magnetic field $\bm B$, which is 
allowed in gyrotropic media only~\cite{Lechner11,Tarasenko_orbital,Tarasenko_orbitalMPGE2}. 
Microscopically, this term is  caused by
structural inversion asymmetry (SIA) and/or bulk inversion asymmetry (BIA).
This process, however, is independent of the spin and the corresponding 
scattering rate for, e.g., SIA, is given by
\begin{equation}
\label{scattering}
\bm{W_{\bm{k} \bm{k}^\prime}} = W_0 + w_{\rm SIA} [\bm{B} \times \left(\bm{k} + \bm{k^\prime} \right)]_z  \:,
\end{equation}
where $W_0$ is the field independent term and $w_{\rm SIA}$ is a measure of the structure inversion asymmetry.
Due to magnetic field dependent scattering, transitions to positive and
negative $k_y^\prime$-states occur with different probabilities.
Therefore hot electrons with opposite $k_y$ have different
relaxation rates in the two spin subbands. In
Fig.~\ref{fig9}(b)  this difference is indicated
by arrows of different thicknesses.
The resulting electric current is given by
\begin{equation}\label{currentorbit}
\bm{j}_{\rm orb} = - 2 e \sum_{\bm{k}} \bm{v}_{\bm{k}} f_{\bm{k}} \:,
\end{equation}
where $\bm{v}_{\bm{k}}=\hbar \bm{k}/m^*$ is the electron velocity and $f_{\bm{k}}$ is the electron distribution function. The latter is found from the Boltzmann equation 
\begin{equation}
G_{\bm{k}} - \sum_{\bm{k}^\prime} \left[\bm{W_{\bm{k} \bm{k}^\prime}} f_{\bm{k}^\prime} (1-f_{\bm{k}}) - W_{\bm{k}^\prime \bm{k}} f_{\bm{k}} (1-f_{\bm{k}^\prime}) \right] =0 \:,
\end{equation}
where the generation term $G_{\bm{k}}$ describes electron gas heating by radiation. 
Since the scattering rate~(\ref{scattering}) contains asymmetric part proportional to $w_{\rm SIA} B$,
the asymmetric part of the distribution function $f_{\bm{k}}$ and, consequently, the photocurrent $\bm{j}_{\rm orb}$
are linearly coupled with the magnetic field and the degree of SIA 
\begin{equation}
\label{orbital}
\bm{j}_{\rm orb} \propto w_{\rm SIA} B\, .
\end{equation}

We note that this dependence remains linear  in the magnetic fields up to~\cite{Tarasenko_orbital,Tarasenko_orbitalMPGE2}
\begin{equation}\label{Borbit}
B \approx \frac{\pi^2\hbar c}{e L_W^2}\:,
\end{equation}
which for  $L_W ~\approx 20$~nm is about 25~T, i.e. much larger than fields used in our experiment.
Here $e$ is electron charge and $c$ is the speed of light.

On the phenomenological level both mechanisms are described by the same 
equations~\cite{Lechner11,Tarasenko_orbital,Tarasenko_orbitalMPGE2} 
and the total current is given by the sum of their
contributions

\begin{equation}
\label{fit}
   j_y = j_{\rm spin} + j_{\rm orb}.
\end{equation}%

Taking into account only the dependence on the magnetic field given by Eqs.~\ref{current1} and ~\ref{orbital}, 
we used for the fitting curves $j_{\rm spin} = a \cdot s(B)$ and $j_{\rm orb} = b \cdot B$, 
where $a$ and $b$ are fitting parameters.
The phenomenological similarity hinders the decomposition of both terms,
because the spin contribution $ J_{\rm spin} \propto  s$
and the orbital one $ J_{\rm orb} \propto  B$ behave identically under a 
variation of the radiation's polarization state and the orientation of the 
magnetic field relative to the crystallographic axes.
Our above consideration shows, however, that the behavior 
of the photocurrent upon a variation of the magnetic field strength 
is different for these  two 
mechanisms.
Combining spin and non-spin mechanisms 
and assuming they have opposing signs 
we can explain the nonlinear magnetic field behavior, 
in particular, the reversal of the photocurrent direction.
Figure~\ref{fig2} shows the results of calculations fitted to the experimental data obtained at 
low power excitation which just slightly 
increase the electron temperature $T_e$ above the lattice temperature $T$, see Fig.~\ref{fig5}(a). 
Using the lattice temperature for calculations and scaling 
$J_{\rm spin}$- and  $J_{\rm orb}$- magnitudes  we obtained  
a good agreement between experiment and the theory in the whole magnetic field range. 
Figures~\ref{fig3} and \ref{fig4} demonstrate that Eq.~(\ref{fit}) 
also describes well the data for the high power excitation 
where the electron temperature is by tens degrees larger 
than the lattice one, see Fig.~\ref{fig5}(b). 
As discussed above, the fact that in these experiments magnetic fields below 7~T and temperatures below 120~K are used, the dependence due to the Zeeman splitting is very weak (see Fig.~\ref{fig8}). Therefore, we obtain good agreement for both lattice temperatures and electron temperatures 
assumed to exceed the lattice temperature by several 
tens of degrees. 
%

Calculations show that these mechanisms yield photocurrents of comparable strength.
At low magnetic fields the total current is dominated by the orbital mechanism. 
However, even at moderate magnetic fields, the nonlinear 
increase of the average spin due to the exchange interaction causes 
an enhancement of the spin-related LMPGE which at high fields becomes 
the major origin. 
The fact that the orbital effect provides a comparable contribution to the spin-related effect is surprising, 
particularly when taking into account that InSb QWs are characterized by the strong
spin-orbit coupling and enhanced magnetic properties. 
Orbital effects, however, are also enhanced in InSb QWs. The reason is the narrow gap  leading to a small effective mass of electrons. 
As demonstrated in Ref.~[\onlinecite{Tarasenko_orbital,Tarasenko_orbitalMPGE2}] the orbital current 
increases with a lower effective mass.

\subsection{Circular MPGE}

The signature of the CPMGE is that the signal is proportional to the radiation helicity and, consequently,
reverses the sign upon switching the helicty from left to right circular polarization, see Fig.~\ref{fig6}. 
In a similar approach to the LMPGE we consider the interplay between the spin and non-spin mechanisms. 
We firstly discuss the spin related contribution which is microscopically due to the 
spin galvanic effect~\cite{Nature02}. 
For the geometry shown in the inset of Fig.~\ref{fig6} 
the magnetic field dependence of the CMPGE 
photocurrent caused by the spin-galvanic effect (see Fig.~\ref{fig10}) 
is given by~\cite{Nature02} 
\begin{equation}
\label{Hanle} J_x \propto -\frac{\omega_L\tau_{s \perp}}{1 + (\omega_L \tau_{s})^2}\:S_{0z}\:,
\end{equation}
where $\tau_s = \sqrt{\tau_{s\parallel} \tau_{s\perp} }$ and
$\tau_{s \parallel}, \tau_{s \perp}$ are the longitudinal and
transverse electron spin relax\-ation times, the Larmor frequency
is given by $\omega_L = g^*\mu_{\rm B} B_x / \hbar$, and $S_{0z}
= \tau_{s\parallel}\dot{S}_z$ is the steady state electron spin
polarization in the absence of a magnetic field. It is seen that the 
photocurrent should follow the well known Hanle law: it achieves the maximum 
of an in-plane spin and consequently the current at 
$\omega_{L}\tau_{s}$ about unity and vanishes for higher magnetic fields. 
The spin relaxation time in our 20~nm InSb-based QW 
has been studied applying circularly polarized pump-probe technique yielding for liquid helium temperature 
$\tau_s \approx 0.1$~ps and $g^* = -45$, see~\cite{Leontiadou}.
Thus, the photocurrent maximum is expected for magnetic fields about 2.5~T. 
In our experiments, however, the current linearly rises 
with the magnetic field and does not exhibit any nonlinearities. 
This fact forces a conclusion that the spin-galvanic effect does not contribute to the CMPGE.

Microscopically, the orbital contribution to the CMPGE appears 
similarly to that of the LMPGE current 
described above~\cite{Tarasenko_orbital,Tarasenko_orbitalMPGE2}. 
The current is caused by the action of the
Lorentz force on the orbital motion of the 
two-dimensional electrons in the radiation field. 
Under irradiation with circularly polarized
light electrons perform a cyclic motion. In the system with SIA/BIA the presence 
of an in-plane magnetic field  pointed along the $[110]$ or $[1\bar{1}0]$ axes forces them to
flow predominantly along the direction of $\bm B$.  Note, that the circular photocurrent, 
sensitive to the radiation helicity sign, is generated due to a retardation between the 
rotating electric field of the radiation and the electron velocity. Therefore, it reaches a 
maximum at $\omega \tau \approx 1$ (here $\omega = 2\pi f$ is the radiation angular frequency 
and $\tau$ is the scattering time) and vanishes for 
much lower or higher frequencies.
The microscopic theory of this effect is given in Ref.~[\onlinecite{Tarasenko_orbital,Tarasenko_orbitalMPGE2}]. 
Like the orbital LMPGE, the resulting orbital current $J_C$  is caused by the 
$B$-dependent corrections to the scattering probability, Eq.~(\ref{scattering}). 
For QWs with $L_W = 20$~nm it is linearly coupled with magnetic field up to $B$ 
about 25~T, see Eq.~(\ref{Borbit}). Thus, on the basis of the magnetic field behavior, 
we conclude that the CMPGE in InSb-based QWs is dominated by the 
orbital mechanism which is in this material enhanced due to the small energy band gap.

\section{Summary}
\label{summary}

Summarizing, our experiments of THz radiation induced 
linear and circular MPGE in InSb-based QW structures show
that due to narrow energy gap, strong magnetic property and 
strong spin-orbit coupling the effect is substantially 
enhanced compared to other III-V materials.
The measurements demonstrate that both spin and orbital 
mechanisms of the MPGE contribute to the signal yielding the current 
contributions of comparable strength.
The observed strong nonlinear behavior of the LMPGE
is caused by nonlinearity of the Zeeman spin splitting and 
supports recent conclusions on the high polarization dependent 
spin susceptibility of a two-dimensional 
electron gas in InSb-based QWs being much larger than 
observed in larger mass systems~\cite{Perez}.

We thank S.A.~Tarasenko, L.E. Golub, and V.~Lechner for fruitful discussions.
Support from DFG (SFB~689), Linkage Grant of IB of BMBF at DLR, and RFBR
is acknowledged. S.K.C. gratefully acknowledges support by EPSRC-UK under grant 
number EP/E055583-1 and A.M.G. the support of EPSRC under grant number EP/F065922/1.


\begin{thebibliography}{99}
\bibitem{symmetry}  In general asymmetric zinc-blende structure based
QWs  grown on  (001) exact oriented substrates 
belong to the $C_{2v}$ symmetry group. A current at $B = 0$ is not
allowed for materials of this symmetry and in fact does not observed in 30~nm QW samples.
Thus, the signal observed in 20~nm QW samples indicates a symmetry reduction 
of the investigated samples due to, e.g., mis-orientation of substrate 
or presence of a strain in the structures. 
We note that although in the 
20~nm QWs structures nonzero signal at $B_{x}$ = 0 
is observed, its
magnitude becomes smaller than that of magnetic field induced 
current even for rather low $B_{x} \approx 0.2$~T.
\bibitem{heating} We note that our estimations show that for 
single pulse excitation the heating of lattice is negligibly small 
and does not contribute to 
the photoconductive signal, 
the result which is well known for such type of measurements~\protect \cite{Ganichevbook}. We also note that 
comparison of photoconductive data with the temperature dependence of mobility 
can not be directly used to determine the electron temperature, because in the second case 
rising of lattice temperature results in increase of number of phonons.
\end{thebibliography}
\end{document}